\documentclass[preprint,12pt,authoryear]{elsarticle}
\usepackage[1.7]{bxpdfver}
\usepackage{here}
\DeclareUnicodeCharacter{3000}{ }

\usepackage{amssymb}
\usepackage{amsmath}

\journal{Nuclear Physics B}

\begin{document}

\begin{frontmatter}


\ead{meistersinger@nifty.com}

\title{Finite Lifetime Fragment Model　4 for Striae Formation in the Dust 
Tails of Comets (FLM 4)
Acceleration by Lorenz-force
} 


\author{Kimihiko Nishioka} 

\affiliation{organization={Hoshinohiroba},
            addressline={LEXCEL GARDEN HACHIOJI 206,
Takakura-machi 39-2
}, 
            city={ Hachioji},
            postcode={192-0033}, 
            state={Tokyo},
            country={Japan}}

\begin{abstract}
The striations in the dust tails of comets are referred to as striae, and their origin has long been a mystery. We introduce a new dynamic model to describe the forms of the striae observed in comets Hale-Bopp (C/1995 O1), West (C/1975 V1), and Seki-Lines (C/1962 C1). Charged particles made of refractory materials, with radii less than 0.5 micrometer, are expelled from the comet's nucleus and accelerated by Lorentz forces near the nucleus. These particles decay many times to form striae, which have a lifespan of less than about 100 days at a distance of 1 astronomical unit from the sun. Over time, they continue to decay and eventually disappear from view. The following dynamic model explains these material science processes. Particles expelled from the comet's nucleus are subjected to three forces: solar gravity, solar radiation pressure, and Lorentz forces near the nucleus. As these particles decrease in size, the Lorentz forces and radiation pressure cause fluctuations, increasing and decreasing to form striae. This model, which is less of a dynamic approximation than previous theories (FLM3), explains the structure of the striae, enables predictions of their luminosity, and clarifies their origin.
\end{abstract}
\begin{highlights}
\item Lorentz force near the nucleus is proposed to explain the form of the striae of comets.
\item The stria is formed by the continuous decay of particles ejected from the comet.
\item We consider that particles above 0.005 $\mu m$ in the process of decay are visible as a stria.
\item Dust particles are subjected to gravity, radiation pressure and Lorentz force.
\end{highlights}
\end{frontmatter}
\section{Introduction}
\label{sec1}

Dust tails of large comets often feature multiple thin strands. These lines, known as striae, have been observed in comets such as Mrkos (C/1957P1), Seki-Lines (C/1962C1), West (C/1975V1) (see Fig. 1), Hale-Bopp (C/1995O1), McNaught (C/2006P1), and Neowise (C/2020F3). The mechanism behind the formation of these striae remains unclear, despite several studies (Nishioka et al., 1992) as shown below.
\begin{figure}[t]
\centering
\includegraphics[bb=0 0 283 213,width=100mm]{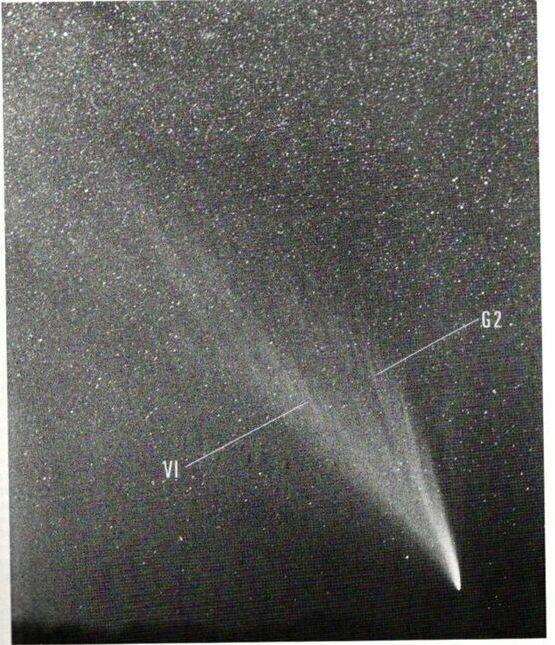}
\caption{Comet West (C/1975 V1) on March 6.810, 1976. Many striae can be seen in the dust tails. The characters indicate the names of striae. Observer: the author (Nishioka, 1998); Lens: f.l. 55mm Fno.\ 2.8; exposure: 8 min; emulsion: Kodak 103aE, without filter.}
\label{fig1}
\end{figure}

\subsection {Sekanina and Farrell's Model of Fragmentation}
\label{subsec1}
Sekanina (1976), Sekanina and Farrell (1980, 1982).
This is the first theory to explain the formation of the striae of the dust tail of the comets.
This model posits that parent particles made of dusts emitted from a comet's nucleus disintegrate en masse to create a stria of the dust tail. This theory requires five independent parameters to match the theory with observed striae, which is relatively few and calculated structure match the observations of comets West (C/1975 V1) and   Mrkos(C/1957 P1). However, there are two major drawbacks: the parent particles forming the striae must all be the same radiation pressure (same size of the parent particles) before decaying, and it is unclear why they break up simultaneously.

\subsection {Finite Lifetime Fragment Model (FLM) (Nishioka and Watanabe 1990)}
\label{subsec2}
A modified version of 1.1, where the parent particle of the dust tail of the striae can break at any time, but the resulting fragments have a lifetime of a few months, continually breaking down until they become invisible. This model can explain the striae of comets West (C/1975 V1), Seki-Lines (C/1962 C1), and Mrkos (C/1957 P1). The issue with this model is that the parent particles forming the stria must be the same radiation pressure (same size of the parent particles) as in 1.1
\subsection { Finite Lifetime Fragment Model 2 (FLM2) (Nishioka 1998)}
\label{subsec3}
This model addresses all the problems of 1.1 and 1.2. The ongoing decay of particles explains the structures of the striae in comets Seki-Lines (C/1962 C1) and West (C/1975 V1), and the striae formation process is clarified by introducing the fragmentation of the complex particles of the unit, made of refractory materials and absorbent materials in terms of material science. However, there are still drawbacks: a dynamical approximation in the theory, eight parameters needed to fit the theory to observations, inability to calculate the brightness of the striae, and the Lorentz force (Price et al. 2019) observed at comet McNaught (C/2006 P1) and comet PanSTARS (C/2014 L4) is not considered.
\subsection {Finite Lifetime Fragment Model 3 (FLM3) (Nishioka 2021)}
\label{subsec4}
This model resolves all issues of 1.1, 1.2, and 1.3. The continuous decay of complex particles explains the striae of comets Seki-Lines (C/1962 C1), West (C/1975 V1), and Hale-Bopp (C/1995 O1). However, it still requires eight parameters to fit the theory to observations, and the Lorentz force (Price et al. 2019, Price et al. 2023) observed at comet McNaught (C/2006 P1) and comet PanSTARS (C/2014 L4) is approximated.
\subsection {Initial Velocity Models (Notni 1964, Notni and Thaenert 1988)}
\label{subsec5}
This model suggests that high-velocity particles ejected from a comet nucleus form a stria. However, it has the disadvantage that observations of the same stria on different days do not align with the theoretically calculated shape.
\subsection {Monte Carlo Methods}
\label{subsec6}
Kharchuk and Korsun (2010) used a Monte Carlo method to simulate the striae of comet C/2006 P1 (McNaught), but the direction of the striae did not match observations.
\subsection {Model of an Optically Dense Dust Mass}
\label{subsec7}
Froehlich and Notni (1988) proposed that striae form due to clumps of optically dense dust ejected from the comet nucleus. However, they did not provide a computational explanation for the observed striae shapes.
\subsection {Fragmentation due to Rotational Acceleration by Sublimation.}
\label{subsec8}
This model assumes that massive ice containing dust, stripped from the comet nucleus, sublimates, accelerates, and rotates. The dusty ice from the cometary nucleus sublimates, accelerates, rotates, and then breaks apart to form a stria (Steckloff and Jacobson 2016). While the distance from the comet nucleus to the stria is estimated, the direction and length of the stria remain unclear.

This theory, or Finite Lifetime Fragment Model 4, represents a significant improvement over FLM3.
\section{Finite Lifetime Fragment Model 4(hereinafter referred to as FLM4)}
\label{sec2}
\subsection {Dynamical Models and the Time - Acceleration Function}
\label{subsec2.1}
	 The ion tail is formed when ions from the comet's coma are swept away by the solar wind magnetic lines. This is examined in detail. Please refer to Figure 2. Magnetic field lines parallel ($\otimes$ in Figure 2) and antiparallel ($\odot$ in Figure 2) to the extension line of the line connecting the sun and the comet nucleus (the central line in Figure 2) are created, and these are connected in a U-shape. At the top of the U shape of B field, current J generated perpendicular to the centerline(according to the right-hand rule of Lenz). Perpendicular to these magnetic lines are the magnetic field B and the flow of ions, which field constitutes the current J (according to the right-hand rule of Lenz), generating the Lorentz force L (according to Fleming's left-hand rule). The relationship is given by $L=J\times B$. Lorentz force L always points to anti-solar direction, and is independent of the azimuth angle around the centerline of the magnetic field B. Because J is always perpendicular to the B. By the Lorentz force L, ions are accelerated to approximately 400 km/s, forming the ion tail (Saito, 1989; Ip, 2004; Ip and Axford, 1982). Similarly, charged silicate particles emitted from the comet nucleus are also accelerated by this Lorentz force, experiencing the acceleration A described by equation (1) (Wallis and Hassan, 1983). But they didn’t mention about striae.
\begin{equation}
A \approx 0.004R/S^2 
\end{equation}
Here, R is a proportional constant (nondimensional), and typical value is 0.4763 (see table1). S is the radius of the particles in micrometers. The unit of A is cm/sec2. The specific gravity of the particles is 3. Below, equation (1) is assumed to also apply to refractory particles. The theory of striae formation according to FLM4 is described next. The ratio of the acceleration A of refractory particles to gravitational acceleration is $\beta_i$
\begin{equation}
\beta_i =6.74527*10^{-3}*R/S^2                  
\end{equation}
It is assumed that A acts only while the particles are within the acceleration cylinder shown in Figure 2. The acceleration cylinder is postulated by the author. It is a virtual space. The radius of the is CL, where the unit of CL is kilometers. The ratio of the solar radiation pressure acting on refractory particles to gravity is:
\begin{equation}
\beta_f=hSi*F(S)
\end{equation}
\begin{equation}
F(S)=0.48*10^{[-1.3\{\log(S/0.24)\}^2]}  
\end{equation}
for S$\le$ 0.6$\mu$m
\begin{equation}
F(S)=0.16*S^{-1.2}  
\end{equation}
for $0.6\mu m\le S$

    Here, F(S) represents the value of radiation pressure relative to gravity for silicate particles, as described by Mukai (1989). The particles have a specific gravity of 2.5. hSi is a proportional constant to fit the theory to the observation. Real particles have various shapes and not spheres. The specific gravity of the particles does not restrict to 2.5. hSi explains, for example, these differences of various $\beta_f^{'s}$.
To summarize, particles ejected from the comet nucleus experience a combined acceleration due to radiation pressure $\beta_f$, Lorentz force $\beta_i$, and solar gravitational acceleration $\beta_g$. $\beta_g$ is -1. $\beta$ is defined by the following equation: 
\begin{equation}	
\beta=\beta_f+\beta_i                            \end{equation}              
Figure 3 illustrates a schematic diagram of the time variation of $\beta$, $\beta_f$, and $\beta_i$. Particles ejected from the comet nucleus follow $\beta_f+\beta_i+\beta_g$ and move along their trajectories, forming striae while disintegrating, as shown in Figure 4. In this dynamical model, particles ejected from the comet's nucleus are assumed to decay multiple times by the sublimation gas pressure of the “glue” of the particle. “Decay” means they become smaller particles. This process is differs from that of (Steckloff and Jacobson 2016). Particles move according to Newtonian mechanics under the influence of the Sun's gravity, solar radiation pressure, and the Lorentz force. The method of numerical calculations is essentially the same as those of (Bessel.,1836; Bredichin.,1886; Finson and Probstein.,1968; Kramer et al 2014) except that the $\beta$ value varies with time. The interplanetary electromagnetic force (Price et al., 2019; Price et al., 2023; Kramer et al., 2014), considered in FLM3 (Nishioka, 2021) as electromagnetic force, is ignored in this model. Because it is smaller than $\beta$.
\newpage
\begin{figure}[H]
\centering
\includegraphics[width=100mm,scale=0.5]{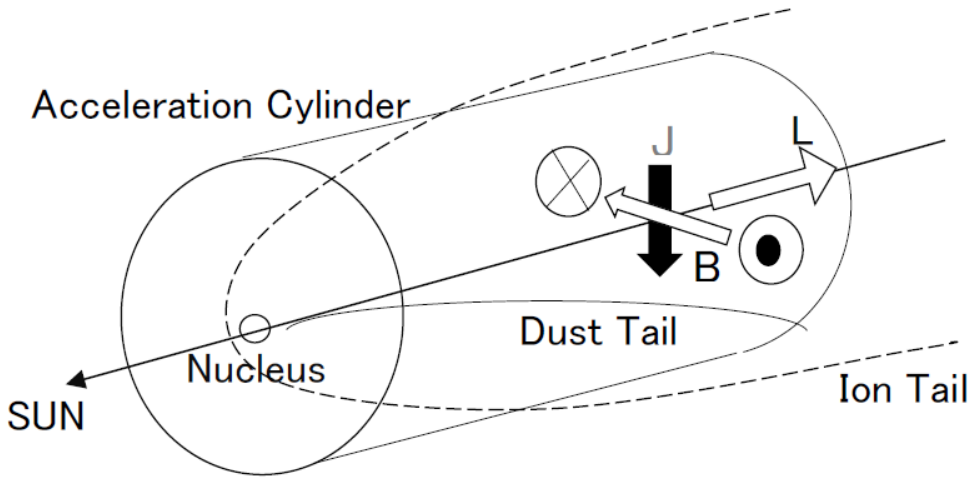}
\caption{Schematic diagram of the magnetic field $B$, current $J$, Lorentz force $L$, ion tail, and dust tail near the comet nucleus.
The line connecting the Sun and the comet nucleus is called the central line. 
The dashed line indicates the ion tail, while the curved solid line indicates the dust tail. 
Magnetic field lines parallel ($\otimes$) and antiparallel ($\odot$) to the central line create a U-shape. 
Perpendicular to these, the magnetic field $B$ and the flow of ions and charged particles (current $J$) generate the Lorentz force $L$. 
The acceleration cylinder has the central line as its center and a radius of CL, with the nucleus at the center on the Sun-facing side. 
Charged particles experience the acceleration $\beta_i$ only when inside the acceleration cylinder. 
This figure is a modification of Saito (1989).}
\label{fig2}
\end{figure}

\begin{figure}[H]
\centering
\includegraphics[scale=0.4]{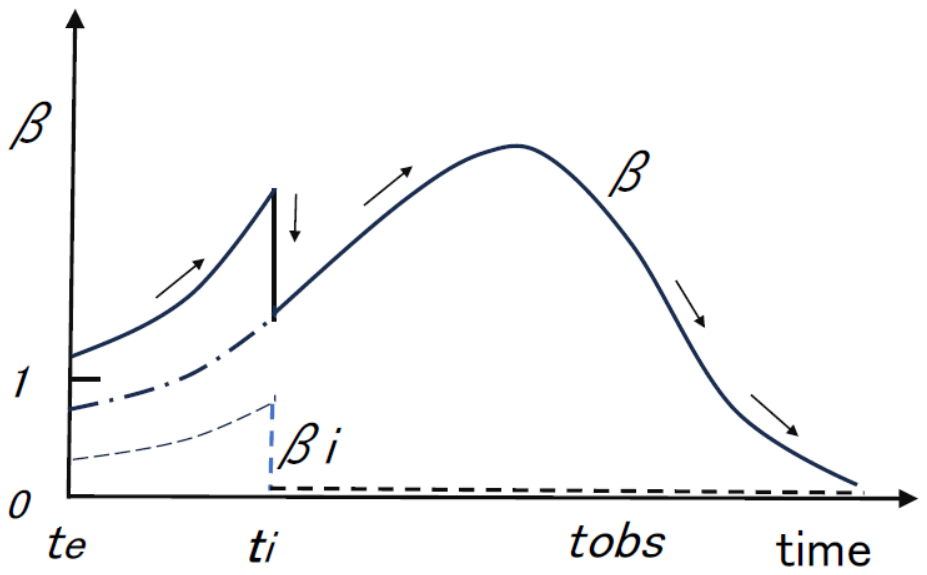}
\caption{Schematic Diagram of the Time Variation of the Ratio $\beta$ of the Acceleration Experienced by Charged Particles to Gravitational Acceleration. (Time- Acceleration function)
The horizontal axis represents time, and the vertical axis represents the ratio of the acceleration experienced by charged particles to gravitational acceleration. The dotted line represents $\beta_f$, the dashed line represents $\beta_i$, and the solid line represents the time variation of $\beta$. ti is the time when the charged particles exit the acceleration cylinder. Particles emitted at time te experience an increase in $\beta_i$ as they disintegrate, reaching zero at ti. After ti, $\beta_i$ remains zero. $\beta_f$ increases after te as the particles disintegrate, peaks, and then decreases. After ti, $\beta$ is equal to $\beta_f$. Therefore, $\beta$ increases after te, decreases at ti, and then rises and falls again. This repeated rise and fall of $\beta$ is the fundamental mechanism for the formation of striae. Small arrows along the graph indicate the time evolution of $\beta$. trajectory of the particles ejected from the nucleus with 0 initial velocity lies on the orbital plane of the comet. The time variation of $\beta$ affects the perihelion distance, the eccentricity and the argument of perihelion of the orbital element of the particle. The mechanism of fragmentation of the particle is out of scope of this study.}
\label{fig3}
\end{figure}
\begin{figure}[H]
\centering
\includegraphics[scale=0.49]{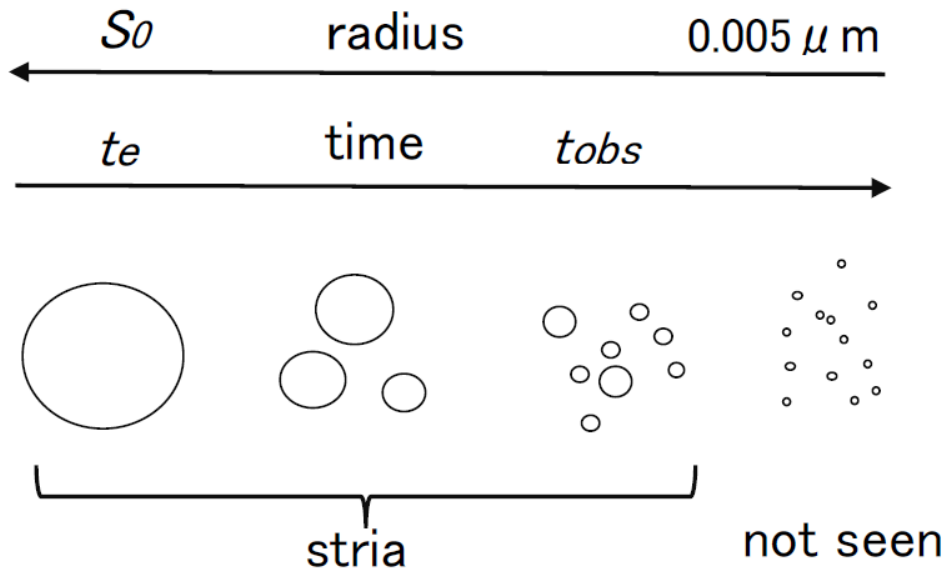}
\caption{ A diagram illustrating the fragmentation process by which particles ejected from a comet's nucleus form a stria.
The horizontal axes depict time (extending to the right) and particle size (extending to the left). Refractory particles ($\bigcirc$) are emitted from the comet's nucleus at time te. These particles break down over time, continuing to disintegrate until they become invisible. According to FLM4, all particles, except those that become invisible, are observed as striae during this process.}
\label{fig4}
\end{figure}
\subsection {Material Science Model}
\label{subsec2.2}
	To elucidate the formation of striae, we propose the following material science model, which describes the process of stria formation as illustrated in Figure 4.
	First, at time te, a particle with radius S0 and zero initial velocity is ejected from the comet's nucleus. The radius of this particle lies within an upper limit S0U and a lower limit  S0L, meaning only particles where $S0L\le S0\le S0U$ are released. Over time, these particles disintegrate due to solar radiation energy, forming smaller particles that create a stria. The fragments are assumed to continuously break down into smaller pieces, eventually becoming invisible when their radius is less than 0.005 $\mu m$, at which point they lose their solid character (Mukai, 1994). This material science model resembles FLM2 (Nishioka, 1998) and FLM3 (Nishioka, 2021), but differs in that complex particles are not required. All refractory particles, except those that become invisible, are observed as striae. Why only particles $S0L\le S0\le S0U$ are released is a problem of the future study. 
	Next, let's describe the decay process dynamically. The value of $\beta_f$ changes as the radius S decreases, as shown in Figure 4. According to Equations (7) and (8), the radius S(t) of a particle is assumed to reduce to 1/e over c days at one astronomical unit from the sun. These equations apply to all particles.
\begin{equation}
S(t)=S0*\exp{-E(t)/c} 
\end{equation}
	
\begin{equation}
E=\int_{te}^{t}\frac{1}{r(S)^2}dt
\end{equation}	
	
	where r(S) is the heliocentric distance of the particle at time t when its radius is S. E is the solar energy received by the particle from the time it was ejected from the comet nucleus up to time t. Equations (7) and (8) imply that the particle's disintegration into smaller particles is proportional to the solar radiation energy impacting the particle per unit area, and this disintegration mechanism occurs when the thickness of the "glue" binding the particles is independent of S0 (Nishioka, 1998).The verification of this  "glue" model has not been carried out.
	
	The details of the fragmentation process are as follows. The radiation pressure of the particle relative to gravity, $\beta_f$ , is a function of the radius S and follows formulas (3), (4), and (5). F(S) represents the value of radiation pressure relative to gravity for silicate particles (Mukai, 1989), and hSi is a proportional constant. For particles where S0 is slightly larger than 0.24 $\mu m$, $\beta_f$ increases with decay and reaches a maximum value $\beta_{fm} = 0.48*hSi$ at $S = 0.24\mu m$. Thereafter, $\beta_f$ decreases. The particles continue to disintegrate, becoming progressively smaller until they are no longer visible. These fluctuations in $\beta$ are the source of stria formation. This decay process differs from that described by Steckloff and Jacobson (2016).
	The FLM4 independent parameters required to reproduce the observed shape of a single stria are the five marked with an asterisk in Table I

\begin{figure}[H]
\centering
\includegraphics[width=230mm,scale=0.8]{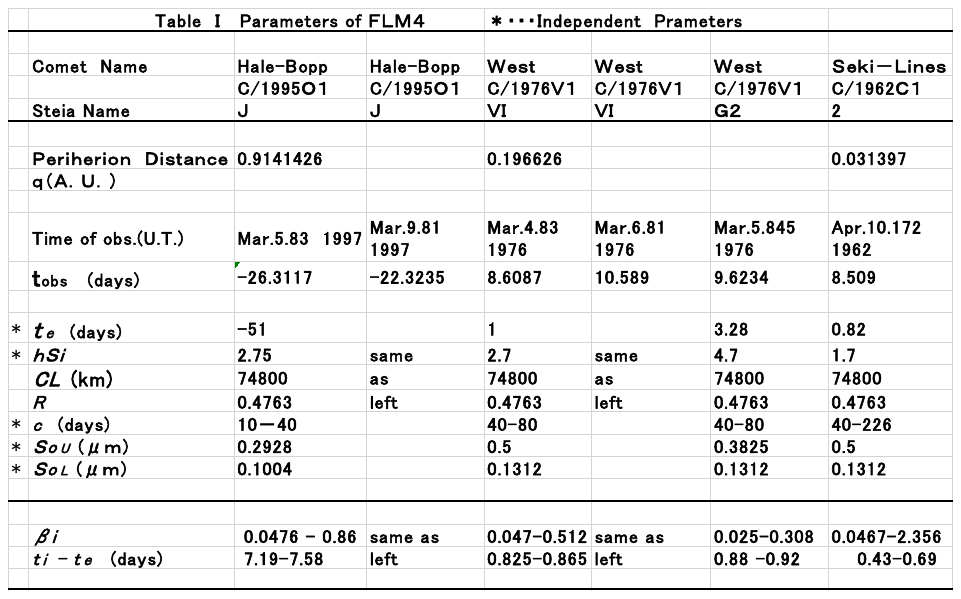}
\caption{Table 1. The asterisk marks denote the independent parameters of FLM4. Units are shown in parentheses. CL and R are constants specific to FLM4 and take nearly identical values for all striae. For comets Hale–Bopp (C/1995 O1) stria J and West (C/1975 V1) stria VI, the same parameters explain observations over a span of two days.}
\label{Table1}
\end{figure}

\section{3.	Comparison of FLM4 with the Observed Striae}
\label{sec3}
A comparison between the observed striae morphologies and those calculated using the dynamical model described in section 2.1, under the following conditions a, b, c, and d, is presented in Figures 5-1 to 5-6.

a. The number of particles expelled from a comet is proportional to $S0^{-3}$

b. The luminosity of the particles forming the stria is calculated, considering Mie scattering and the solar spectrum.

c. The area of the square symbols in Figures 5-1,5-2,5-3,5-4,5-5,5-6 is approximately proportional to the brightness of the particles.

d. All refractory particles and their fragments are visible as striae.
\begin{figure}[H]
\centering
\includegraphics[scale=0.5]{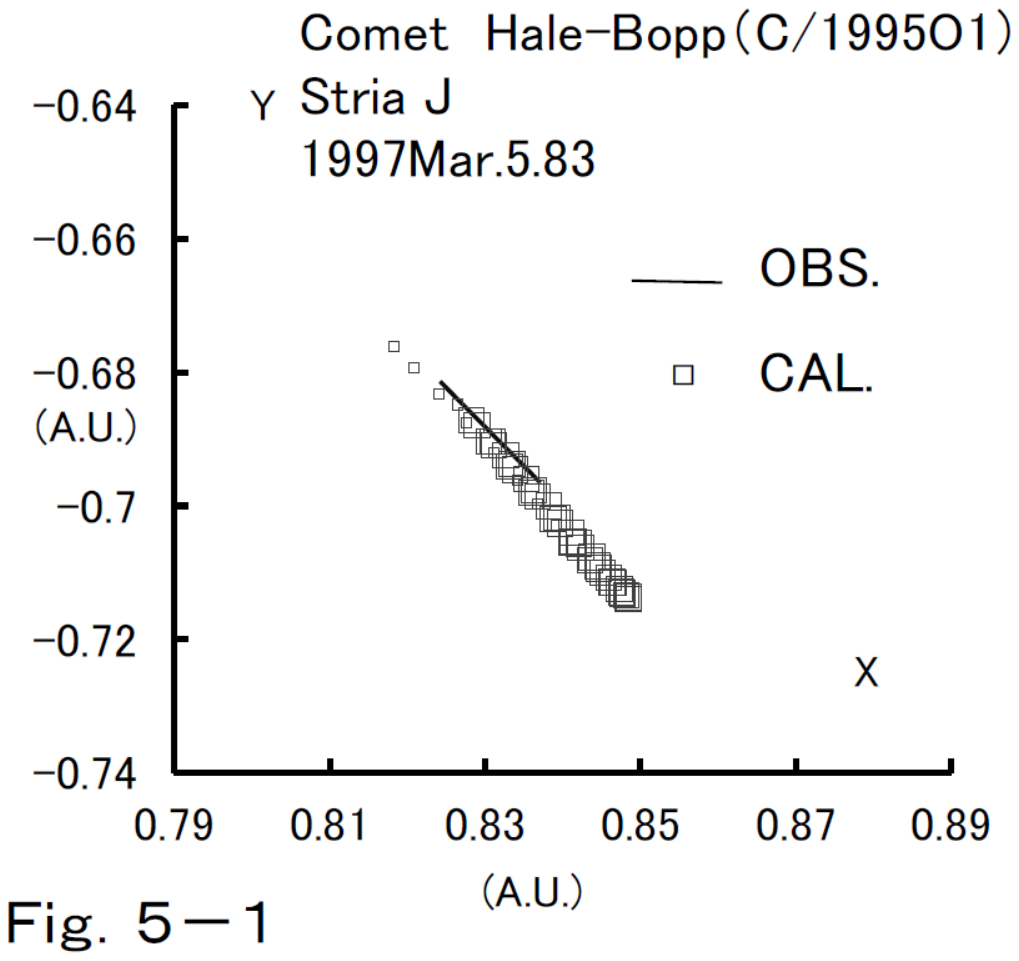}
\caption{Model calculation result for stria segment 5–1.}
\label{fig5-1}
\end{figure}

\begin{figure}[H]
\centering
\includegraphics[scale=0.5]{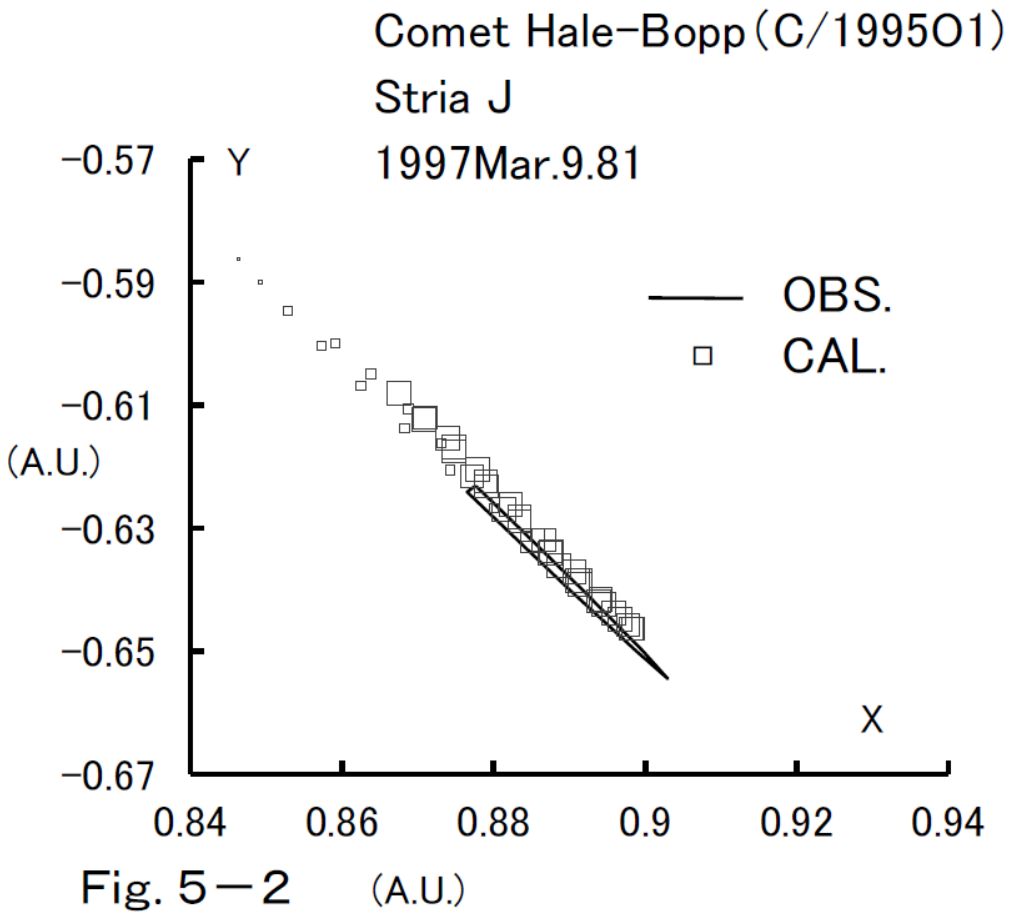}
\caption{Model calculation result for stria segment 5–2.}
\label{fig5-2}
\end{figure}

\begin{figure}[H]
\centering
\includegraphics[scale=0.5]{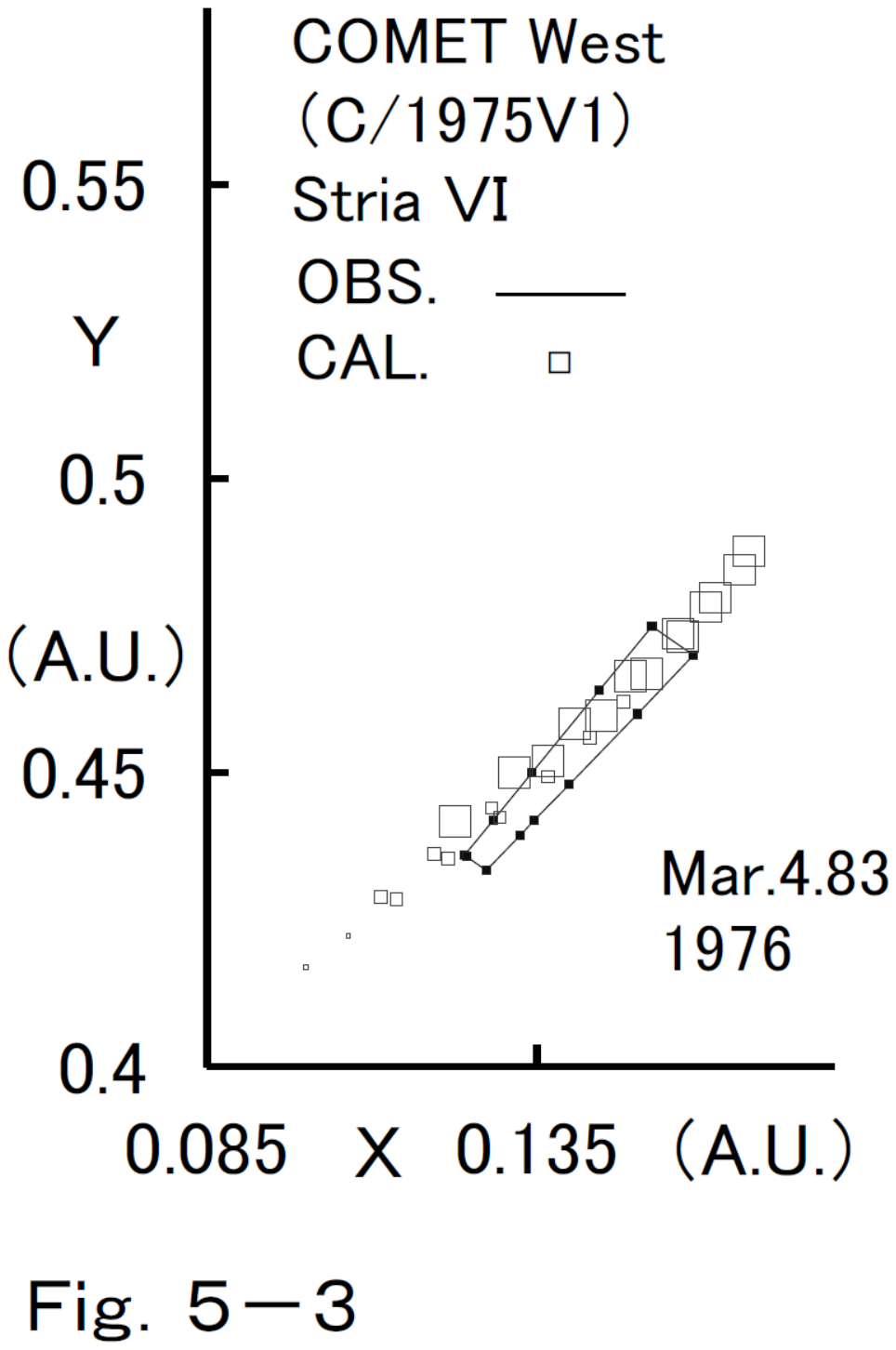}
\caption{Model calculation result for stria segment 5–3.}
\label{fig5-3}
\end{figure}

\begin{figure}[H]
\centering
\includegraphics[scale=0.5]{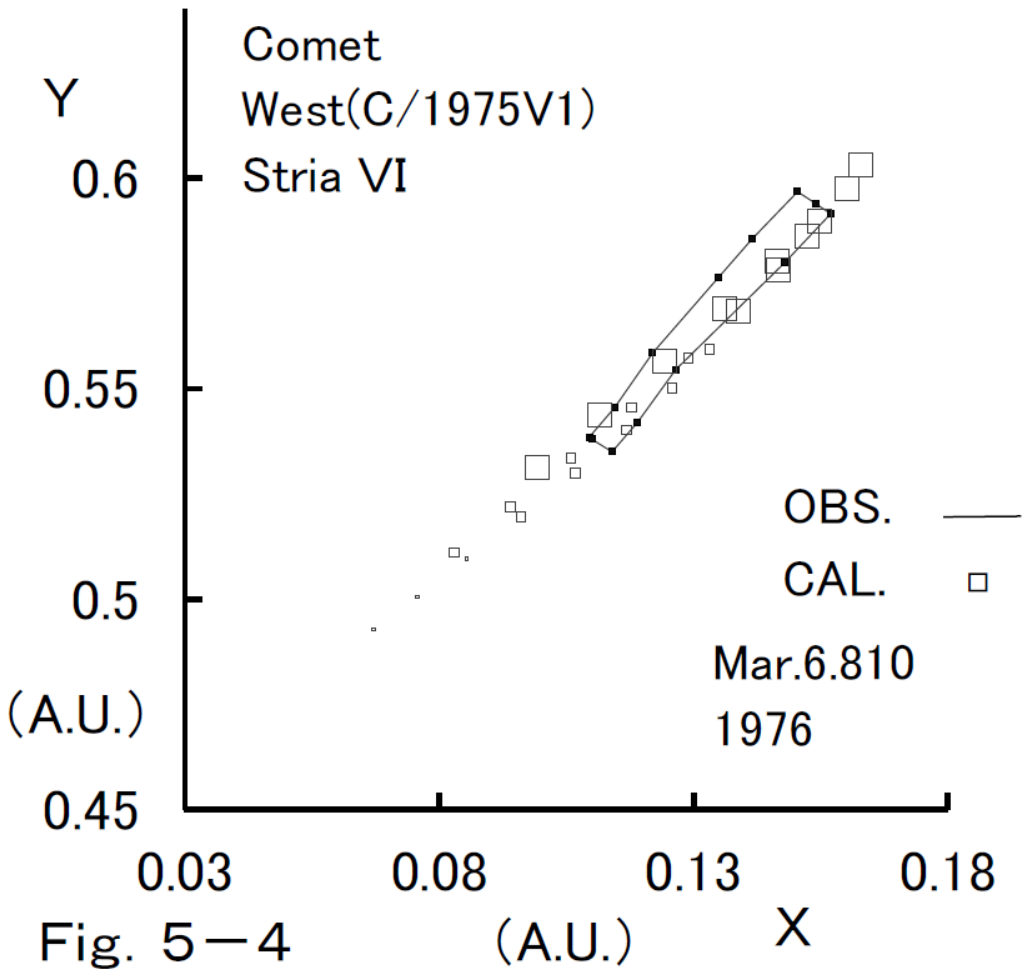}
\caption{Model calculation result for stria segment 5–4.}
\label{fig5-4}
\end{figure}

\begin{figure}[H]
\centering
\includegraphics[scale=0.5]{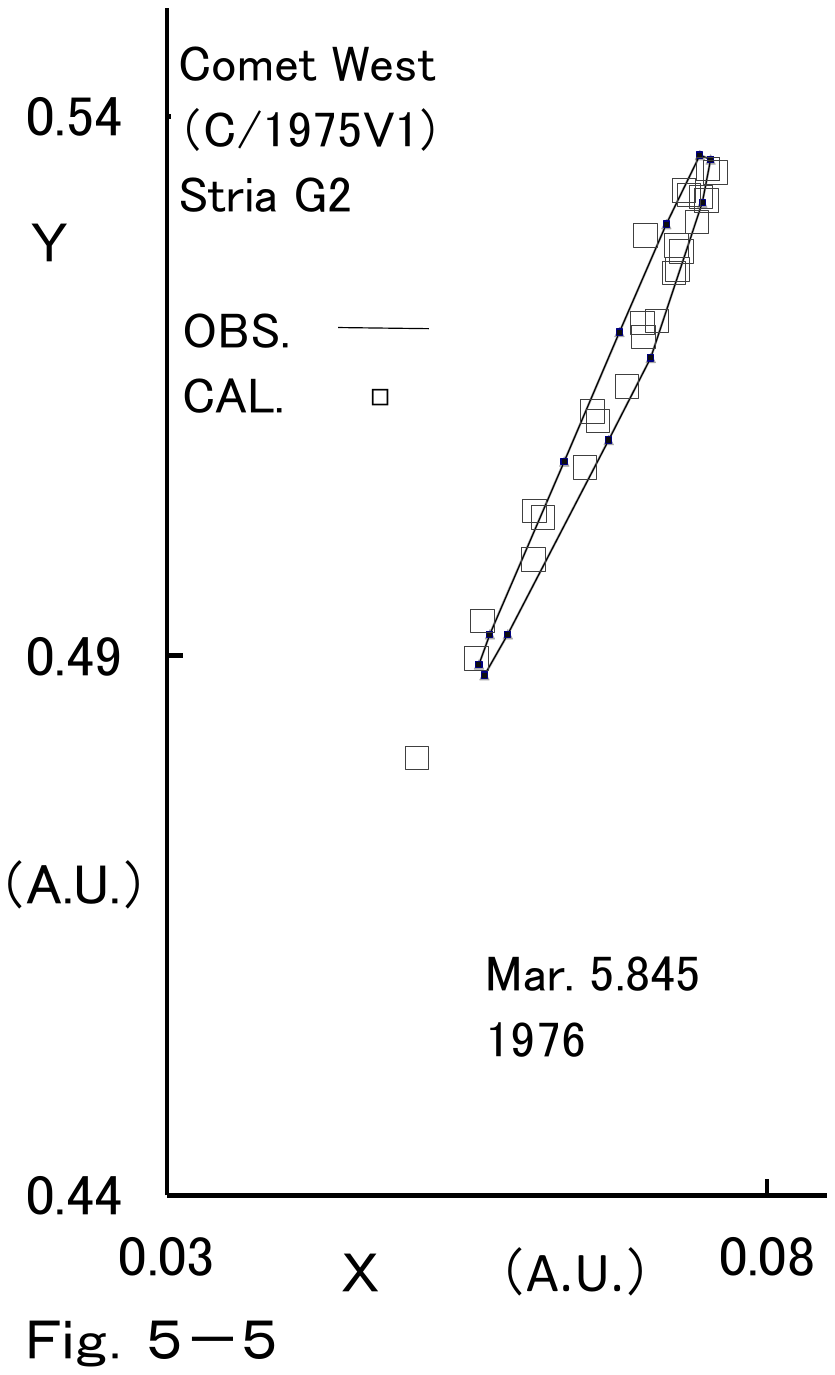}
\caption{Model calculation result for stria segment 5–5.}
\label{fig5-5}
\end{figure}

\begin{figure}[H]
\centering
\includegraphics[scale=0.5]{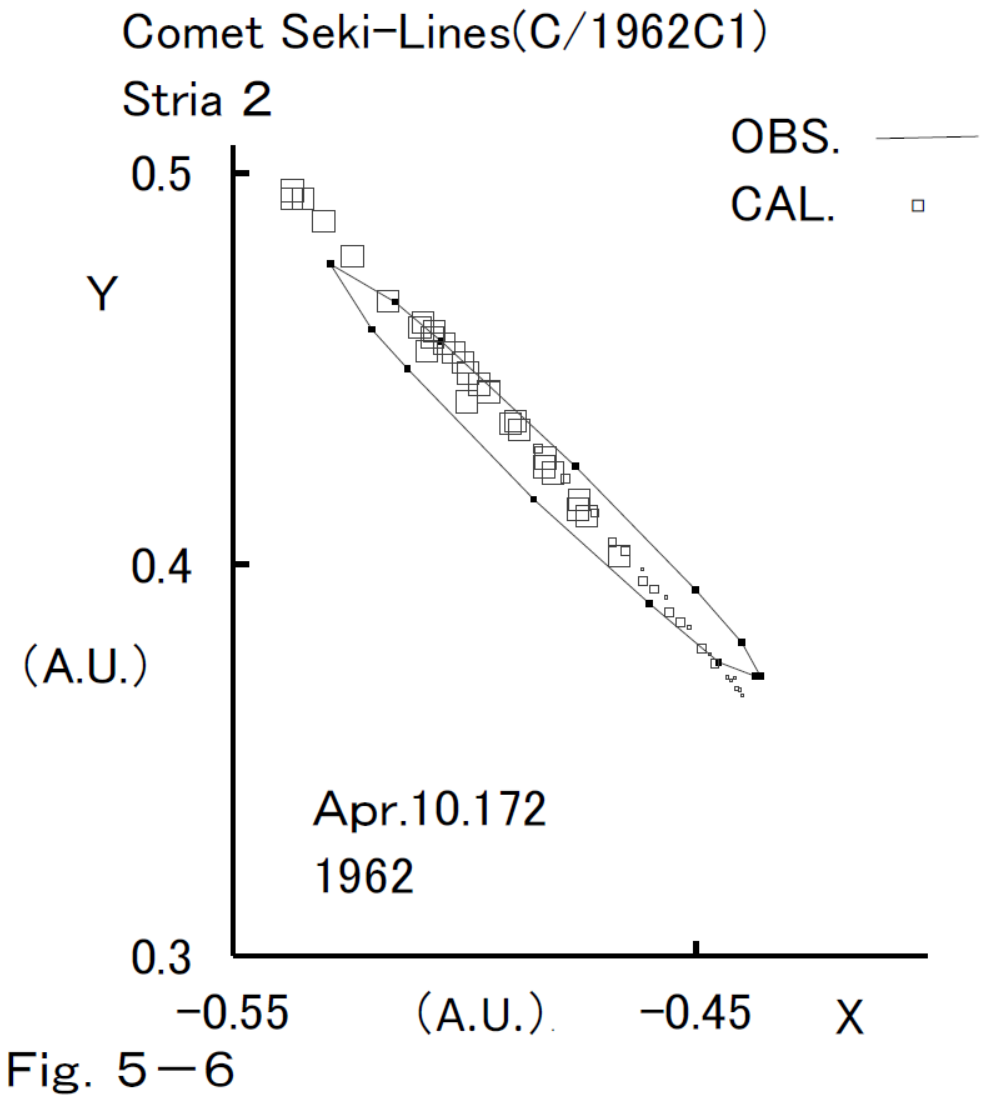}
\caption{Model calculation result for stria segment 5–6.}
\label{fig5-6}
\end{figure}

Figure 5. Comparison of observed striae structures with theoretical calculations Using FLM4

Figures 5-1 through 5-6 display diagrams in the orbital plane of the comet, where the sun is positioned at the origin and the direction of the perihelion is on the +X axis. The +Y axis shows the direction of the comet after it has passed its perihelion. The unit of measurement is in astronomical units (A.U.). The observed shape(outline) of the stria is represented by a polygonal solid line as stira has width, and is projected onto the orbital plane(OBS.). The calculated positions of the stria particles at tobs using FLM4 are shown as squares. The size of each square is roughly proportional to the particle's luminosity. The row of squares represent the stria(CAL.). Particles of approximately three different c’s values and ten different S0’ s values were ejected from the nucleus at te . Ultimately, around 30 particles were released from the comet nucleus, and their positions are illustrated.
Figures 5-1 and 5-2 compare the theoretical model with observations of comet Hale-Bopp (C/1995O1) stria J over 4-day intervals. Figures 5-3 and 5-4 compare the theoretical model with observations of comet West (C/1975V1) stria VI over 2-day intervals. Figures 5-5 and 5-6 compare the theoretical model with observations of comet West (C/1975V1) stria G2 and comet Seki-Lines (C/1962C1) stria 2, respectively. Refer to Table I for the parameters used in the calculations.
The particle luminosity was computed using the same method as in FLM3 (Nishioka 2021, See section 3, Eq17,18,19, and 20).

	As demonstrated in Figs. 5-1 to 5-6, the stria morphologies calculated by FLM4 for three comets with perihelion distances differing by factors of ten roughly match with the observations. The same parameters were used to describe the observed morphologies of comets Hale-Bopp (C/1995O1) stria J and West (C/1975V1) stria VI over two days. The parameter values for this model FLM4, corresponding to the calculated values in Figures 5-1 to 5-6, are listed in Table I. This table shows that CL and R are theoretically specific constants for any stria. 
	In case of the Fig5-1. and Fig5-2. calculations do not match the observations in lengths direction of the striae. This reason may be the identifications of stria J of two days wrong. In case of the Fig5-3. And Fig5-4. calculations roughly match the observations. 
	Inclinations of the striae are a little different. In case of the Fig5-5. And Fig5-6. calculations very well match the observation.
	
	Photographic data used for the striae observations are as same as table II of FLM3 (Nishioka, 2021).

\section{4. Discussion}
\label{sec4}
The formation of striae requires the acceleration of particles emitted from the comet nucleus to increase, decrease, increase, and decrease (Nishioka, 2021). To achieve this, FLM3 introduced the continuous disintegration of composite particles for the formation of striae (Nishioka, 2021). FLM4 successfully explained the formation of striae by introducing the Lorentz force near the comet nucleus as the cause of the acceleration's increase and decrease, combined with the continuous disintegration of charged refractive particles. The latter is a more natural theory than the former. The parameter R in equation (1) is proportional to the voltage (in volts) of the charged particles, with R = 0.4763, a natural value. While CL is somewhat large, doubling the value of CL and halving the value of R still explained the shape of the striae. Similarly, reducing CL and increasing R could also explain the shape of the striae. R and CL shared common values across six striae. Treating these as independent variables and optimizing them for each stria could further improve the match between theory and observation. It is assumed that CL and R do not depend on the heliocentric distance of the comet. The length of the acceleration cylinder is about ten times CL. The values of $\beta i$ range from 0.025 to 2.356, which are reasonable. If hSi = 3, $\beta fm$ is 1.44, which is also reasonable.
The flow of the solar wind is assumed to be in a steady state. If there are disturbances in the solar wind, the direction of the acceleration cylinder, CL, and R would change, and the shape of the striae would also bend, curve, or change. Bent striae have been observed in the comet Seki-Lines (C/1962C1) (McClure, 1962; Nishioka and Watanabe, 1990). Refractive materials were chosen as the material for the charged particles, because using absorbent materials would result in discrepancies between the observed striae and theoretical values. Table I shows the values of the five independent parameters, the two constants CL and R, and $\beta i$ and ti - te. ti - te  is the time it takes for particles emitted from the comet nucleus to exit the acceleration cylinder.
	Striae are formed when the $\beta$ of the particles expelled from the comet nucleus changes with the decrease in S. $\beta$ changes according to the Fig.3, increasing, decreasing, increasing and decreasing. These up’s and down’s of $\beta$ are the fundamental mechanism for the formation of striae. In section 2.2, we supposed that composition of the particles are refractory particles. For example, materials such as basalt, obsidian, andesite, and quartz are also candidates for the composition of the particles. Mixture of afore mentioned materials are also allowed.

\section{5. Conclusion}
\label{sec5}

5.1 FLM4 has shown that the striae in the dust tail are formed through a mechanism similar to that of the ion tail, with the common factor being the acceleration caused by the Lorentz force acting on charged particles near the nucleus.

5.2 Striae form as the radius of particles emitted from the comet nucleus decreases, leading to fluctuations in $\beta$—increasing, decreasing, increasing, and then decreasing. These variations in acceleration are the fundamental cause of striae formation, a concept also present in FLM3.

5.3 FLM4 successfully accounts for the shapes and brightness of the six striae observed in three different comets.

5.4 In FLM4, only five parameters are required to explain the shape of a single stria, with CL and R effectively being constants. This is three fewer than the eight parameters needed by FLM3, making FLM4 a more straightforward theory.

5.5 In FLM4, the particles emitted from the comet nucleus do not need to be composite particles as in FLM3. It is assumed that all refractory particles and their fragments can form striae, making FLM4 a more natural and simplified theory compared to FLM3.

Acknowledgement
I am thankful to Prof. Junichi Watanabe, Mr. Hideo Fukushima, Mr. Daisuke Kinoshita, Mr. Ken Sugawara, and Mr. Masayuki Takata for providing the positional data of stria J. I also extend my gratitude to amateur astronomers Yukihisa Kato, Shirou Sugimoto, and Alan McClure for supplying the photographic materials of the striae.







\end{document}